\documentclass[aps,twocolumn,groupedaddress,floatfix]{revtex4}
\usepackage{graphics}%

\begin{document}

\bibliographystyle{apsrev}

\title{Electromagnetically induced transparency and reduced
speeds for single photons in a fully quantized model}

\author{Thomas Purdy}
\affiliation{Department of Physics, Carnegie Mellon University, Pittsburgh,
PA 15213}
\author{Martin Ligare}
\email[]{mligare@bucknell.edu}
\affiliation{Department of Physics, Bucknell University, Lewisburg, PA 17837}

\date{\today}

\begin{abstract}
We introduce a simple model for electromagnetically induced
transparency in which all fields are treated quantum mechanically. We
study a system of three separated atoms at fixed positions in a
one-dimensional multimode optical cavity. The first atom serves as the
source for a single spontaneously emitted photon; the photon scatters
from a three-level $\Lambda$-configuration atom which interacts with a
single-mode field coupling two of the atomic levels; the third atom serves
as a detector of the total transmitted field. We find an analytical
solution for the quantum dynamics. From the quantum amplitude
describing the excitation of the detector atom we extract information
that provides exact single-photon analogs to wave delays
predicted by semi-classical theories.  We also find complementary
information in the expectation value of the electric field intensity
operator.
\end{abstract}


\maketitle

\section{Introduction}

Controlling the phase coherence in ensembles of multilevel atoms has
led to the observation of many striking phenomena in the propagation
of near-resonant light.  These phenomena include coherent population
trapping, lasing without inversion, electromagnetically induced
transparency, and anomalously slow and anomalously fast pulse
velocities.  Resonant enhancement of the index of refraction without
accompanying increased in absorption was proposed \cite{SCU91} and
observed \cite{BOL91,FIE91} in 1991, and drastic reductions in the
group velocity of pulses were discussed shortly thereafter
\cite{HAR92}.  Recent experiments have taken the reduction of the
speed of light to extreme limits \cite{HAU99,KAS99} and at the other
extreme lie observations of seemingly superluminal light
\cite{WAN00,DOG01}.  An overview of recent developments in the control
of photons is presented in Ref.~\cite{LUK01}.  An earlier review of
electromagnetically induced transparency was presented by Harris
\cite{HAR97}, while Lukin et al. present a overview of phase coherence
in general, with an extensive list of references \cite{LUK99}. Such
effects are also discussed in a recent texts (see, for example,
Ref.~\cite{SCU97}). In most previous work the phenomenon of
electromagnetically induced transparency and the accompanying drastic
slowing of the speed of light are treated using semi-classical theory
in which the atoms of the medium are treated quantum mechanically and
the fields are treated classically. We use a model in which the entire
system is treated quantum mechanically, and study the propagation of a
field state containing a single photon.  Although coherent states of a
single mode quantized field are often considered as the ``most
classical,'' the single photon states that we study exhibit striking
parallels with classical fields.

We study a system of three separated atoms at fixed positions in a
one-dimensional multimode optical cavity. The first atom serves as the
source for a single spontaneously emitted photon; the photon scatters
from a three-level $\Lambda$-configuration atom which interacts with
an additional single-mode field coupling two of the atomic levels; and
the third atom serves as a detector of the total transmitted field. In
the spirit of Feynman's derivation of the classical index of
refraction from the interaction of a field with a single oscillator
\cite{FEY63}, we infer the properties of a medium exhibiting
electromagnetically induced transparency from the interaction of the
spontaneously emitted quantum field with the single quantized
scattering atom.  We find an analytical solution for the quantum
dynamics, including reradiation from the scatterer, and from this we
deduce quantum delays that characterize the propagation of the field.
These delays are equivalent to those predicted by semi-classical
theories.  In our quantum model all delays are clearly the result of
interfering amplitudes that reshape the temporal envelope of the
probability of detecting the transmitted photon.  This effect is most
clearly illustrated in the graphs of detection probability vs.  time
displayed in Sec.~\ref{sect_cog_delay}.  This work is an extension of
the model we have used previously to study quantum manifestations of
classical wave delays induced by scattering from simple two-level
atoms \cite{PUR02}.  We note that the analytical results obtained in
this paper may be verified using the straightforward numerical
techniques like those used in Refs.~\cite{TAY01,LIG01a,LIG02,BUZ99}.

\section{Review of semi-classical theory}
\label{sc_sec}
\begin{figure}[t]
\includegraphics{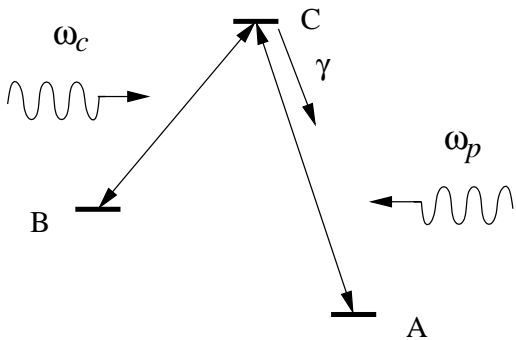}
\caption{Level scheme for simple electromagnetically induced 
transparency.  A strong field resonantly couples levels B and C, 
and  weak probe is near resonance with the transition between 
levels A and C.  The upper level C decays via spontaneous emission
at a rate $\gamma$ to the ground state A.  All other damping 
mechanisms are assumed to be negligible.}
\label{f_level1}
\end{figure}

Electromagnetically induced transparency can be observed in the simple
three-level atom illustrated in Fig.~\ref{f_level1}.  A strong laser
``coupling'' laser with angular frequency $\omega_c$ is tuned to
resonance with the transition between atomic levels B and C, while a
weak ``probe'' laser with angular frequency $\omega_p$ excites the
transition between levels A and C.  In this paper we consider the
simplest case in which decay from level C to B is small enough to be
neglected, and the only damping is due to emission at a rate $\gamma$
from level C to the ground state A.  (In the this paper all decay
rates $\gamma_j$ refer to decay of {\em probability} of finding the
atom in the excited state.)  We also assume that there is no
incoherent pumping driving population between the levels of the atom.

\begin{figure}[b]
\includegraphics[175,470][300,680]{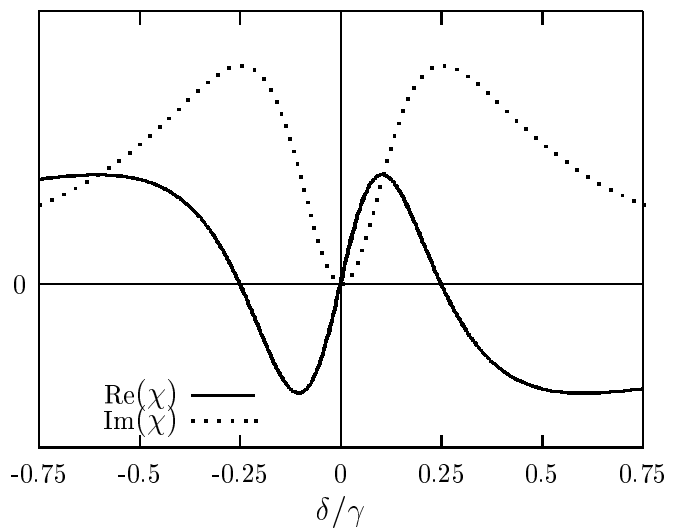}
\caption{Real and imaginary parts of the complex susceptibility for 
the three-level system illustrated in Fig.~\ref{f_level1}.  The strength
of the coupling field is such that $\omega_R=\gamma/4$.}
\label{f_lineshape}
\end{figure}

In the limit of a weak probe, the complex susceptibility is given 
by \cite{SCU97}
\begin{equation}
\chi = \left(\frac{N\vert d\vert^2}{\hbar\epsilon_0}\right)
	\frac{\delta}{\omega_R^2/4 - \delta^2 - i\delta\gamma/2},
\label{chi_eq}
\end{equation}
where $N$ is the density of the atoms, $d$ is the dipole moment of the
transition between levels A and C, $\omega_R$ is the Rabi frequency of
the coupling transition between levels B and C, and the detuning of
the probe laser frequency from resonance is
\begin{equation}
\delta = \omega_p - \omega_{AC}.
\end{equation}
The real and imaginary parts of the susceptibility are illustrated
in Fig.~\ref{f_lineshape} for a coupling field strength such 
that  $\omega_R=\gamma/2$.
The index of refraction and the absorption coefficient can be
calculated from the real and imaginary parts of the complex
susceptibility respectively.  When the probe field is resonant with
the transition between levels A and C, i.e., $\delta=0$, the
absorption goes to zero and the index of refraction is a rapidly
varying function of probe frequency.

For later comparison with the results of our quantum model, we
consider a classical monochromatic plane wave of frequency $\omega$
which is normally incident on a thin slab containing atoms with the
level structure illustrated in Fig.~\ref{f_level1}.  The plane of the
slab is normal to the $z$ axis, the thickness of the slab is $\Delta
z$, and the density of the atoms is $N$.  If the incident field 
is $E_i=E_0e^{-i\omega(t-z/c)}$, the transmitted field on the 
far side of the slab is 
\begin{equation}
E_t = E_ie^{i\omega \chi\Delta z/(2c)},
\end{equation}
and for weak scattering the transmitted field is approximately
\begin{eqnarray}
E_t &\simeq& E_i\left(1 +  i\frac{\omega\Delta z}{2c}\chi\right) \nonumber \\
	&=& E_i \left[1 + i\left(\frac{N\Delta z\vert d\vert^2\omega}
		{\hbar\epsilon_0c\gamma}\right)
	\frac{\delta\gamma/2}{\omega_R^2/4-\delta^2 - i\delta\gamma/2}\right].
		\label{weak_scattering_eq}
\end{eqnarray}
The scattering is characterized by the 
dimensionless parameter $N\Delta z\vert d\vert^2\omega/
(\hbar\epsilon_0c\gamma)$.

\begin{figure}[t]
\includegraphics[175,470][300,680]{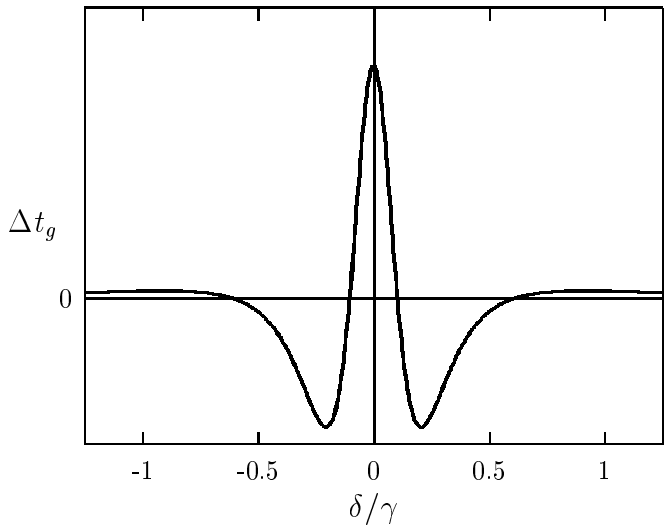}
\caption{Group delay for a classical pulse in a medium with
susceptibility given by Eq.~(\ref{chi_eq}).  As in
Fig.~\ref{f_lineshape}, the strength of the coupling field is such
that $\omega_R=\gamma/2$.}
\label{f_delay}
\end{figure}

The dispersion in the response of the atoms in this model leads to
delays of pulses traversing such a slab.  The delay in the arrival
of the peak of a modulation envelope of a quasi-monochromatic pulse
that is away from a region of anomalous dispersion is determined by
the group velocity $v_g=d\omega/dk=c/(n +\omega\frac{dn} {d\omega})$,
and is given by
\begin{eqnarray} 
\Delta t_g&=& \frac{\Delta z}{v_g}-\frac{\Delta z}{c} \nonumber \\
          &=& \frac{\Delta z}{c}\left(n-1 + \omega \frac{dn}{d\omega}\right).
\end{eqnarray}	   
For detunings such that $\delta\ll\omega_{AC}$ the group delay is
\begin{eqnarray}
\Delta t_g &=& \left(\frac{2N\Delta z\vert d\vert^2\omega_{AC}}
		{\hbar\epsilon_0c}\right) 	\nonumber \\
	&& \times\frac{(\omega_R^2+4\delta^2)\left[(\omega_R^2-4\delta^2)^2 
			- 4\gamma^2\delta^2\right]}		
	{\left[(\omega_R^2-4\delta^2)^2 + 4\gamma^2 \delta^2\right]^2}.
				\label{group_delay_eq}
\end{eqnarray}
The functional form of the delay when $\omega_R=\gamma/2$ is illustrated 
in Fig.~\ref{f_delay}.  
When the probe field is resonant with the the transition between
levels A and C, not only is the absorption zero, but
the group velocity can be extremely small, and this is manifested in
the large positive group delay at $\delta=0$ in Fig.~\ref{f_delay}.
The central peak in Fig.~\ref{f_delay} becomes taller and narrower as
the strength of the coupling field is reduced.  We note that for some
values of the detuning $\delta$ the group delay is negative, which
corresponds to group velocities greater than the vacuum speed of light
$c$.  Such ``superluminal'' velocities do not violate causality, and
are an effect of pulse reshaping by the dispersive medium.  Similar
reshaping effects and group velocities greater than $c$ occur near
simple two-level resonances in both classical and quantum theories
\cite{PUR02}.

For pulses that are either not sufficiently monochromatic, or not
sufficiently far away from a region of anomalous dispersion, the
simple concepts of phase and group velocity are inadequate to
characterize all of the effects of pulse-reshaping as a field
propagates.  Several other velocities and delays have been developed
(see, for example, Refs.~\cite{SMI70,BLO77}) and in this paper we
focus on a delay determined by the ``temporal center of gravity'' of
the field intensity of a pulse at a fixed position $z$ ``downstream''
from the slab containing the atoms comprising the medium, i.e.,
\begin{eqnarray}
\Delta t_{{\cal E}^2} &=& \left(\frac{\int t {\cal E}(z,t)^2\, dt}
                        {\int  {\cal E}(z,t)^2\, dt}
                                \right)_{\mbox{after medium}} \nonumber \\ 
                        && - \left(\frac{\int t {\cal E}(z,t)^2\, dt}
                        {\int  {\cal E}(z,t)^2\, dt}
                \right)_{\mbox{vacuum}}.  \label{cl_tcom_delay_eq}
\end{eqnarray}
This is closely related to concepts used to define the centrovelocity
in \cite{SMI70}.  We have investigated this delay in classical and
quantum mechanical models of scattering from simple two-level atoms in
a previous paper \cite{PUR02}.  For quasi-monochromatic pulses far
from resonance this delay is equivalent to the group delay, but in
general it is necessary to calculate explicitly the field $\cal{E}$ in
order to determine $\Delta t_{{\cal E}^2}$.  For the specific fields
considered in this paper we will show that the
``temporal-center-of-gravity'' delay happens to be equal to twice the
group delay.

\section{Quantum Mechanical Model}

\begin{figure}[b]
\includegraphics{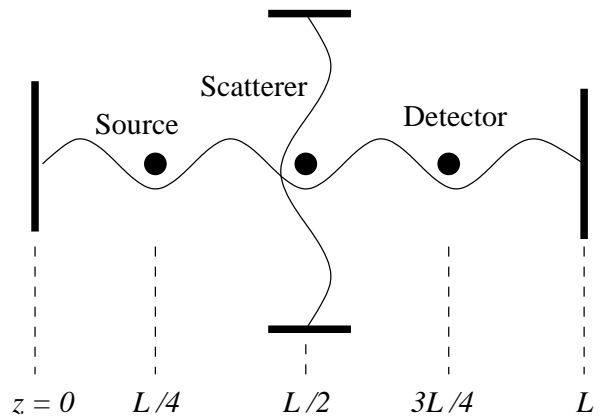}
\caption{Quantum mechanical model consisting of three atoms at 
fixed positions.  All three atoms interact with the radiation in the 
probe field in the {\em multimode} cavity (represented horizontally) and 
the middle atom also interacts with the single mode coupling field
(represented vertically).}
\label{f_model}
\end{figure}

The quantum mechanical system we consider is illustrated in
Fig.~\ref{f_model}, and consists of three atoms in a multimode
one-dimensional optical cavity that extends from $z=0$ to $z=L$.  This
multimode cavity is oriented horizontally in the schematic
representation of Fig.~\ref{f_model}.  The middle atom has an
additional interaction with a single-mode field contained in the
vertical cavity.  The field in the multimode (horizontal) cavity plays
the role of the probe field, and the field in the single mode
(vertical) cavity represents the coupling field.  (The finite optical
cavities do not contribute to the physical phenomena under
investigation; they simply provide a convenient quantization volume
for the field modes used in our calculation.) In the remainder of this
section we will discuss the details of our model and the standard
quantum optical Hamiltonian we use.  We also present the analytical
solution for the time dependence of the system.

\begin{figure}[t]
\includegraphics{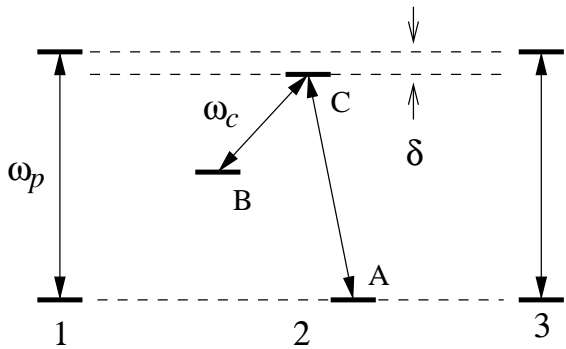}
\caption{Level scheme of the source, scattering, and detector atoms.
The source and detector are two-level atoms with identical transition
frequencies.  The scattering atom is a three-level atom in the $\Lambda$ 
configuration.  The transition frequency of the source and 
detector are detuned from the A-C transition of the scattering atom by 
an amount $\delta$.}
\label{f_level2}
\end{figure}

The atom on the left (atom 1) is a two-level atom which is initially
in the excited state, and will be the source of the probe field.  The
middle atom (atom 2) which will scatter the radiation emitted by the
source is a three-level atom with the ``lambda'' configuration of
Fig.~\ref{f_level1}, with the energy difference between levels A and C
close to that of the level separation of atom 1.  (The highest level
of all three atoms will be labeled as C.)  Levels B and C of atom 2
will interact with the single-mode coupling field which is assumed to
be exactly on resonance.  The coupling field will initially be in a
state with a well-defined number of photons such that the Rabi
frequency of the transition between levels B and C is appropriate for
the observation of electromagnetically induced transparency. 
The two-level atom on the right (atom 3) will serve as a detector.  The
detector atom is assumed to have the same resonant frequency
as the source atom.  The relative energy levels of all three
atoms are illustrated in Fig.~\ref{f_level2}.

We assume that the coupling strengths of the atoms to the probe field
are all independent, so that the atoms have separate spontaneous decay
rates.  Although we find a solution for the dynamics for any decay
rates, we focus our attention in this paper on cases in which the
decay rate of atom 1 is very much smaller than the decay rate of atom
2.  This condition insures that the spectrum of the radiation emitted
by atom 1 will be very much narrower than the linewidth of the the
scattering atom, and this allows us to compare our results from the
quantum case with those of the semi-classical model which assumes that
the sample is driven by a monochromatic wave.  We also assume that the
decay rate of the detector atom (atom 3) is very much greater than any
other dynamical rates in the problem.  In this limit the excitation of
atom 3 will closely follow the field drives it.  For simplicity we
also assume that the spontaneous decay rate from
level C to level B (for atom 2) is negligible.  
  
The zero-field resonance frequencies of the A-C transitions of the
atoms are labeled $\omega^{\rm (at)}_j$, where $j=1$, 2, or 3, and the
positions of the atoms will be labeled $z_j$.  In the remainder of the
paper we will assume that the atoms are at positions $z_1=L/4$,
$z_2=L/2$, and $z_3=3L/4$, as illustrated in Fig.~\ref{f_model},
although our results for delay times do not depend on the exact
positions.  The standing wave field modes of the probe field cavity are 
separated in angular frequency by the fundamental frequency
\begin{equation}
\Delta = \pi\frac{c}{L}.
\end{equation}
This mode spacing may be small enough that many modes fall within the
natural line-width of the atoms.  

For convenience we assume that the frequency of one of the modes
corresponds exactly to the resonance frequency of atom 1, the emitting
atom, and that the length of the cavity is such that it contains an
even number of wavelengths of this mode. We label the frequency of
this mode $\omega_0=k_0\Delta$, where $k_0$ is an integer divisible
by 4.  (This assumption affects the details of some of our
calculations, but not our results concerning delay times.) The other
mode frequencies will be enumerated from this mode so that
\begin{equation}
\omega_k= (k_0+k)\Delta,
\end{equation}
where $k=0,\pm 1,\pm 2,\dots$

As in the semi-classical case we wish to study the effects of the
detuning of the source field on the scattering of the radiation.  We
use the same symbol $\delta$ as in the semi-classical case to represent the
detuning of the field, but in the quantum case the detuning is
directly tied to the properties of the source and scattering atoms:
\begin{equation}
\delta = \omega_1^{\rm (at)} - \omega_2^{\rm (at)}.
\end{equation}
Because the detector atom is assumed to have the same resonance frequency 
as the source atom we have $\omega^{\rm (at)}_1 = \omega^{\rm (at)}_3$.

We use as basis states the eigenstates of the atomic plus free-field
Hamiltonian
\begin{widetext}
\begin{eqnarray}
\hat{H}_0 &=& \hat{H}_{\rm atoms} + \hat{H}_{\rm field} \nonumber \\
	  &=& \hat{H}_{\rm atoms} + \hat{H}_{\rm probe} + 
		\hat{H}_{\rm coupling} \nonumber \\
          &=& \hbar\omega^{\rm (at)}_1\vert C_1\rangle\langle C_1\vert 
	  +\hbar\omega^{\rm (at)}_2\vert C_2\rangle\langle C_2\vert
	  +\hbar(\omega^{\rm (at)}_2-\omega_c)\vert B_2\rangle\langle B_2\vert 
	+ \hbar\omega^{\rm (at)}_3\vert C_3\rangle\langle C_3\vert
	+ \sum_k \hbar\omega_k a^\dagger_k a_k +
		\hbar\omega_c a^\dagger_c a_c,
\end{eqnarray}
\end{widetext}
where $a_k$ and $a^\dagger_k$ are the lowering and raising operators
for the $k^{\rm th}$ mode of the probe field, and $a_{\rm c}$ and
$a_{\rm c}^\dagger$ act similarly on the single mode of the coupling
field.  (We have re-zeroed the energy scale to remove zero-point
energy of the field modes.) The basis states will be denoted as
follows:
\begin{itemize}
\item $\vert C,A,A;\emptyset, N\rangle$ --- Atom 1 excited,
atoms 2 and 3 in the ground state, no photons in the cavity modes,
$N$ photons in the coupling mode;
\item $\vert A,C,A;\emptyset, N\rangle$ --- Atom 2 in state C, atoms 
1  and 3 in the ground state, no photons in the probe field,
$N$ photons in the coupling mode;
\item $\vert A,A,C;\emptyset,N \rangle$ --- Atom 3  excited, atoms 
1 and 2 in the ground state, no photons in the probe field,
$N$ photons in the coupling mode;
\item $\vert A,A,A;1_k,N\rangle$ --- All atoms in the ground state, one 
photon in the probe field mode with frequency $(k_0 + k)\Delta$, $N$ 
photons in the coupling mode; 
\item $\vert A,B,A;\emptyset,N+1\rangle$ --- Atom 2 in state B, atoms
1 and 3 in the ground state, $N+1$ photons in the coupling field.
\end{itemize}

Each atom is coupled to all of the standing wave modes of the probe field,
and we use the symbol $g_{jk}$ to label the coefficients characterizing
the strength of the coupling of the $j^{\rm th}$ atom to the $k^{\rm th}$
mode of the probe field.  In addition atom 2 interacts with the coupling 
field, and this interaction is characterized by the constant $g_c$.
We use the standard electric-dipole and rotating-wave approximation
approximations to give the following interaction Hamiltonian:
\begin{eqnarray}
\hat{H}_{\rm int} &=& \sum_{j=1}^3 \sum_k \hbar \left(
	g_{jk}a_k^\dagger\vert A_j\rangle\langle C_j\vert + 
	g_{jk}^\ast a_k\vert C_j\rangle\langle A_j\vert \right) 
					\nonumber \\
	 && + \hbar \left(g_{\rm c} a_{\rm c}^\dagger 
		\vert B_2\rangle\langle C_2\vert + \Omega_{\rm c}^\ast 
			a_{\rm c} \vert C_2 \rangle\langle B_2 \vert\right).
\end{eqnarray} 
The Rabi frequency $\omega_R$ of the transition between levels B and C
is determined by the number of photons in the coupling mode and the
coupling constant $g_c$ as follows:
\begin{equation}
\omega_R = 2 g_c \sqrt{N+1}.
\end{equation}
For convenience we assume that $\omega_R$ is real.

We assume that the frequencies of all atomic transitions are very much
greater than the fundamental frequency of the cavity, that is
$\omega^{\rm (at)}_j\gg \Delta$ for all atoms, and
similarly for $\omega_c$.  In this limit we can make the approximation
that all modes that influence the dynamics of the system are near the
atomic resonances, and the atom-field coupling constants can be
factored into a product of a frequency-independent constant and a
space-dependent coupling factor.  The coupling constants $g_{jk}$ are
given in terms of the electric dipole matrix element $d_j$ between the
levels A and C of atom $j$, the effective volume of the cavity, $V$,
and the permittivity of free space, $\epsilon_0$, by
\begin{eqnarray}
g_{jk} &=& d_j\left(\frac{\omega^{\rm (at)}_j}{2\hbar\epsilon_0 V}\right)
                \sin\left[(k_0+k)\pi z_j/L\right] \nonumber \\
       &=& \Omega_j\sin\left[(k_0+k)\pi z_j/L\right],
\end{eqnarray}
where in the last line we have defined the quantity 
\begin{equation}
\Omega_j = d_j\left(\frac{\omega^{\rm (at)}_j}{2\hbar\epsilon_0 V}\right),
\end{equation}
which is independent of the cavity mode-frequency.

\begin{figure}[t]
\includegraphics[6.5cm,7.7cm][10cm,24cm]{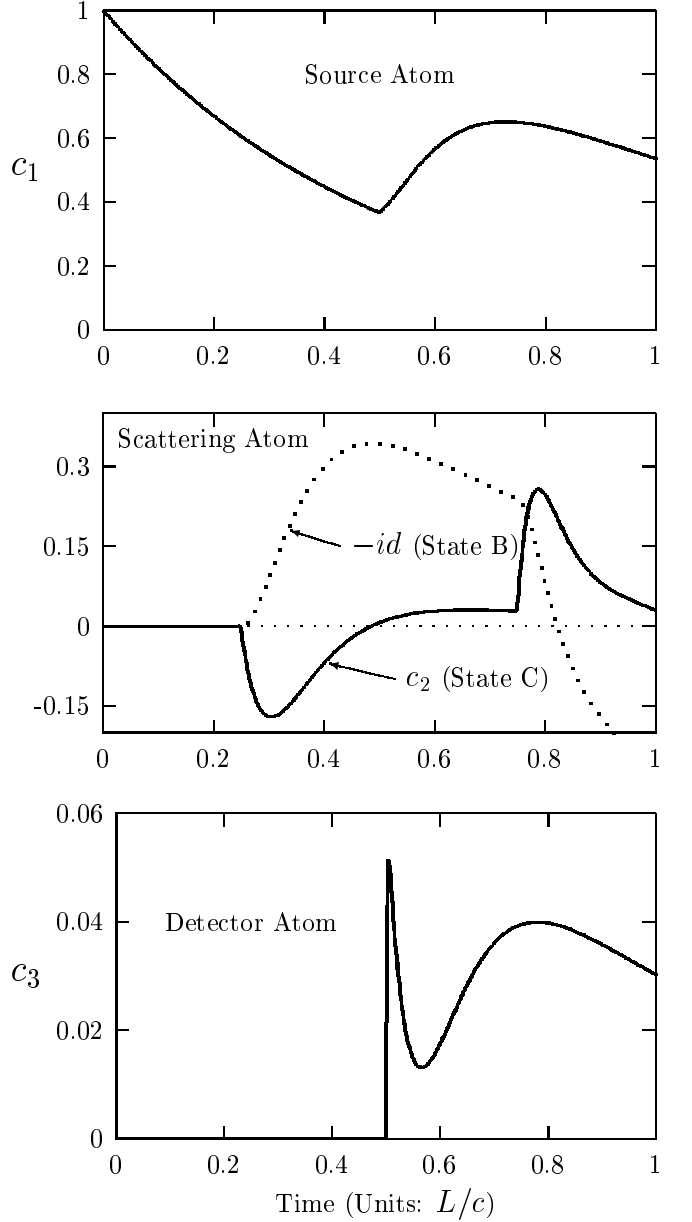}
\caption{Magnitude of the amplitudes for the atoms to be in the 
excited state, starting from the state $\vert\psi(0)\rangle = 
\vert e,g,g;\emptyset,N\rangle$. The decay rates of the atoms are 
$\gamma_1 = 4$,$\gamma_2 = 64$, and $\gamma_3 = 1024$; the Rabi frequency
of the B-C transition of atom 2 is $\omega_R = \gamma_2/2 = 32$, and 
$\delta = 0$.}
\label{f_sol2}
\end{figure}

We write the state of the system as the linear combination
\begin{eqnarray}
\vert\psi(t)\rangle &=& c_1(t)\vert C,A,A;\emptyset, N\rangle 
			+ c_2(t)\vert A,C,A;\emptyset, N\rangle \nonumber \\
		&&	+ c_3(t)\vert A,A,C;\emptyset, N\rangle \nonumber \\
		&&      + d(t)\vert A,B,A;\emptyset, N+1\rangle
							\nonumber \\
		&&      + \sum_k b_k(t)\vert A,A,A;1_k,N\rangle.
\label{gen_psi_eq}
\end{eqnarray}

Choosing the zero of the energy scale at the level of (uncoupled)
state $\vert C,A,A;\emptyset, N\rangle$, the Schr\"{o}dinger yields
the following set of coupled differential equations:
\begin{eqnarray}
\dot{c}_1 &=& -i\sum_k g_{1k}b_k, \label{c1_schro_eq} \\
\dot{c}_2 &=& -i\left(\sum_k g_{2k}b_k + \omega_R d/2 - \delta c_2\right),\\
\dot{c}_3 &=& -i\sum_k g_{3k} b_k, \\
\dot{d}   &=& -i(\omega_R c_2/2 - \delta d), \\
\dot{b}_k &=& -i\left(\sum_j g_{jk}^\ast c_j + k\Delta b_k\right).
		\label{bk_schro_eq}
\end{eqnarray}

We solve this set of equations with the Laplace transform technique
used by Stey and Gibberd \cite{STE72} for a Hamiltonian similar to
that for a single atom at the center of one-dimensional cavity.
Laplace transforms have also been used to solve the Schr\"{o}dinger
equation in similar problems with two interacting atoms in
three-dimensions \cite{MIL74,MIL75} and in our own recent work on
scattering from two-level atoms \cite{PUR02}.  Because the Laplace
transform technique is not new, and because we would like to focus on
analogies with the fields in the semi-classical model and physical
interpretation, we leave the details of our solution to the appendix,
and simply quote our results here.

The general features of the solution giving the time-dependences of
the atomic excitation amplitudes are illustrated in Fig.~\ref{f_sol2}.
The initially excited atom decays exponentially until the time
$t=0.5L/c$ at which scattered and reflected radiation first returns to
the atom.  The amplitudes to find the other atoms excited are
identically zero until radiation first reaches them: the scattering
atom first becomes excited at $t=0.25L/c$ and the detector atom is
first excited at $t=0.5L/c$.  The three decay constants which
characterize the spontaneous emission rates of level C in each of the 
atoms emerge naturally in terms of the parameters of the Hamiltonian as
\begin{equation}
\gamma_k = \frac{\pi\vert\Omega_k\vert^2}{\Delta} = 
\vert\Omega_k\vert^2\frac{L}{c}.
\label{gamma_def_eq}
\end{equation}
The causal nature of the dynamics is evident in that all disturbances
are propagated at the speed of light $c$ via the quantum field. The
abrupt changes in the complex amplitudes at intervals of $0.5L/c$ are
a manifestation of the finite speed of light and the atomic spacing of
$0.25L$.

The abrupt changes appear in our analytic solution for the
complex amplitudes $c_j(t)$, $d(t)$, and $b_k(t)$ as sums of terms with step
functions that ``turn on'' at successively later intervals of
$0.5L/c$.  In the following formulas giving these amplitudes we
truncate the expressions so that only the first excitation of atoms 2
and 3 are included.  We also note that the following equations are
specific in some details to the atomic positions $z_i$ used in this
paper. The ``turn-on'' times and relative phases of terms will change
with different positions, but the conclusions of this paper concerning
delay times are unaffected by these details.  The time
dependence of the system is given by the following set of amplitudes,
in which we use the labels ${\cal F}_i$ as a shorthand for complicated factors
that are functions of $\gamma_i$, $\delta$, $\omega_R$, and (in the case
of $b_k$), $k$:
\begin{widetext}
\begin{eqnarray}
c_1(t) &=& e^{-\frac{\gamma_1}{2}t} + \Theta\left(t-\frac{1}{2}\right)
	\left\{ e^{-\frac{\gamma_1}{2}(t-\frac{1}{2})}	
	\left[{\cal F}_1 + \left(t - \frac{1}{2}\right){\cal F}_2 \right]
			\right. \nonumber \\
	&& \left.
	  + e^{[-(\gamma_2-\sqrt{\gamma_2^2-4\omega_R^2})/4
		+i\delta](t-\frac{1}{2})}{\cal F}_3 
	   +        e^{[-(\gamma_2+\sqrt{\gamma_2^2-4\omega_R^2})/4
		+i\delta](t-\frac{1}{2})}{\cal F}_4
	\right\} + \cdots \label{c1_gen_eq} \\	
\nonumber \\
c_2(t) &=& \Theta\left(t-\frac{1}{4}\right)
	\left\{ e^{-\frac{\gamma_1}{2}(t-\frac{1}{4})}{\cal F}_5
	  + e^{[-(\gamma_2-\sqrt{\gamma_2^2-4\omega_R^2})/4
		+i\delta](t-\frac{1}{4})}{\cal F}_6 
	   +        e^{[-(\gamma_2+\sqrt{\gamma_2^2-4\omega_R^2})/4
		+i\delta](t-\frac{1}{4})}{\cal F}_7
	\right\} + \cdots \\
\nonumber \\
c_3(t) &=& \Theta\left(t-\frac{1}{2}\right)
	\left\{ e^{-\frac{\gamma_1}{2}(t-\frac{1}{2})}{\cal F}_8
	  + e^{[-(\gamma_2-\sqrt{\gamma_2^2-4\omega_R^2})/4
		+i\delta](t-\frac{1}{2})}{\cal F}_9 
	   +        e^{[-(\gamma_2+\sqrt{\gamma_2^2-4\omega_R^2})/4
		+i\delta](t-\frac{1}{2})}{\cal F}_{10} \right. \nonumber \\
	    && \left. + e^{-\frac{\gamma_3}{2}(t-\frac{1}{2})}{\cal F}_{11}
			\right\} + \cdots \label{c3_gen_eq} \\
\nonumber \\
d(t) &=& \Theta\left(t-\frac{1}{4}\right)
	\left\{ e^{-\frac{\gamma_1}{2}(t-\frac{1}{4})}{\cal F}_{12}
	  + e^{[-(\gamma_2-\sqrt{\gamma_2^2-4\omega_R^2})/4
		+i\delta](t-\frac{1}{4})}{\cal F}_{13} 
	   +        e^{[-(\gamma_2+\sqrt{\gamma_2^2-4\omega_R^2})/4
		+i\delta](t-\frac{1}{4})}{\cal F}_{14}
	\right\} + \cdots \\
\nonumber \\
b_k(t) &=& \left(e^{-\frac{\gamma_1}{2}t}-e^{-ik\pi t}\right)
					g_{1k}{\cal F}_{15}
	+ \Theta\left(t-\frac{1}{4}\right)g_{2k}
	   \left\{e^{-\frac{\gamma_1}{2}(t-\frac{1}{4})}{\cal F}_{16} 
		+ e^{-ik\pi(t-\frac{1}{4})}{\cal F}_{17} \right. \nonumber \\
		&& \left.
                +   e^{[-(\gamma_2 -\sqrt{\gamma_2^2 - 4\omega_R^2})/4
		+i\delta](t-\frac{1}{4})}{\cal F}_{18} 
	   + e^{[-(\gamma_2+\sqrt{\gamma_2^2-4\omega_R^2})/4
		+i\delta](t-\frac{1}{4})}{\cal F}_{19}
	\right\} + \cdots
	\label{bk_gen_eq}
\end{eqnarray}
\end{widetext}
Complete expressions for the factors ${\cal F}_i$ may be found in the 
appendix.

Four time scales are manifest in the exponential factors in these
these equations: the exponential decay rate of the source atom
$\gamma_1$, which is assumed to be small, the exponential decay rate
of the detector atom $\gamma_3$, which is assumed to be large, and two
decay rates associated with the scattering atom,
$\left(\gamma_2+\sqrt{\gamma_2^2-4\omega_R^2}\right)/2$ and
$\left(\gamma_2-\sqrt{\gamma_2^2-4\omega_R^2}\right)/2$.  When the
coupling field strength is small, the last of these rates approaches
zero.  This means that decay from the scattering atom can be very
slow, and this slow reradiation is at the origin of the reduced
velocities which characterize media with electromagnetically induced
transparency.

In the following sections we will focus on two quantities: $c_3(t)$,
the amplitude to find the detector atom excited, and $\langle\hat{\cal
E}^2\rangle$ the expectation value of the square of the electric field
operator, which is proportional to the field intensity.  (The
expectation value of the field operator itself is $0$ for any
single-photon state.)  In our investigation of $c_3(t)$ we will
consider only the first term in Eq.~(\ref{c3_gen_eq}) describing the
initial excitation of the detector atom.  Similarly, we will
investigate $\langle\hat{\cal E}^2\rangle$ in regions to the right the
scattering atom, and at times that exclude multiple scattering
effects.

It is useful to rewrite $c_3(t)$ as the sum of two pieces, the
amplitude $c^0_3(t)$ for atom 3 to be excited in the absence of the
scattering atom (or, equivalently, when $\gamma_2=0$), plus $c_3^s(t)$, 
the amplitude that is attributable to scattering:
\begin{equation}
c_3(t)\equiv c_3^0(t) + c_3^s(t).  \label{divide_c_eq}
\end{equation}
Setting $\gamma_2=0$ in Eq.~(\ref{c3_gen_eq}) gives
\begin{eqnarray}
c^{0}_3(t) &=& \Theta\left(t-\frac{1}{2}\right)
        \frac{\sqrt{\gamma_1\gamma_3}}{\gamma_1-\gamma_3} \nonumber \\
        &&  \times\left(e^{-\frac{\gamma_1}{2}(t-\frac{1}{2})} -
                e^{-\frac{\gamma_3}{2}(t-\frac{1}{2})} \right), \label{c30_eq}
\end{eqnarray}
and subtracting this from Eq.~(\ref{c3_gen_eq}) gives
\begin{widetext} 
\begin{eqnarray}
c_3^s(t) &=& \Theta\left(t-\frac{1}{2}\right)
		\left\{ e^{-\frac{\gamma_1}{2}(t-\frac{1}{2})}{\cal F}_{20}
	  + e^{[-(\gamma_2-\sqrt{\gamma_2^2-4\omega_R^2})/4
		+i\delta](t-\frac{1}{2})}{\cal F}_{21} 
	   +        e^{[-(\gamma_2+\sqrt{\gamma_2^2-4\omega_R^2})/4
		+i\delta](t-\frac{1}{2})}{\cal F}_{22} \right. \nonumber \\
		&& \left. + 
		e^{-\frac{\gamma_3}{2}(t-\frac{1}{2})}{\cal F}_{23}\right\} 
		        \label{c3s_eq}
\end{eqnarray}
where 
\begin{eqnarray}
{\cal F}_{20}&=& \left(\frac{\sqrt{\gamma_1\gamma_3}}{\gamma_1-\gamma_3}\right)
		\frac{\gamma_2(\gamma_1 + i 2\delta)}{
		 \gamma_1^2 - \gamma_1\gamma_2 - 4\delta^2 +  \omega_R^2
		+ i2\delta(2\gamma_1-\gamma_2)}  \label{f20_eq} \\
{\cal F}_{21}&=& \left(\frac{2\sqrt{\gamma_1\gamma_3}}
			{\sqrt{\gamma_2^2-4\omega_R^2}}\right)
	    \frac{\gamma_2\left(\gamma_2-\sqrt{\gamma_2^2-4\omega_R^2}\right)}
	   {\left(2\gamma_1-\gamma_2+\sqrt{\gamma_2^2-4\omega_R^2}+ 
						i4\delta\right)
	    \left(2\gamma_3-\gamma_2+\sqrt{\gamma_2^2-4\omega_R^2}+ 
						i4\delta\right)}\\
{\cal F}_{22}&=& \left(\frac{2\sqrt{\gamma_1\gamma_3}}
		{\sqrt{\gamma_2^2-4\omega_R^2}}\right)
	    \frac{\gamma_2\left(\gamma_2+ \sqrt{\gamma_2^2-4\omega_R^2}\right)}
		{\left(2\gamma_1-\gamma_2-\sqrt{\gamma_2^2-4\omega_R^2}+ 
				i4\delta\right)
	      \left(2\gamma_3-\gamma_2-\sqrt{\gamma_2^2-4\omega_R^2}+ 
			i4\delta\right)}\\
{\cal F}_{23}&=& \left(\frac{\sqrt{\gamma_1\gamma_3}}{\gamma_1-\gamma_3}\right)
		\frac{\gamma_2(\gamma_3 + i 2\delta)}{\gamma_3^2 
		- \gamma_2\gamma_3 - 4\delta^2 +\omega_R^2 
		+ i2\delta(2\gamma_3-\gamma_2)} \label{f23_eq}
\end{eqnarray}

In our investigations of the quantum field itself we use the electric field 
operator in the form given by Meystre and Sargent \cite{MEY99}, and 
write the expectation value of the square of the field as
\begin{equation}
\langle \hat{\cal{E}}^2\rangle = \langle \psi(t) \vert \left\{ \sum_k
        \sqrt{\frac{\hbar\omega_k}{\epsilon_0 V}} \left(a_k +
        a_k^\dagger\right) \sin\left[(k_0 + k)\frac{\pi z}{L}\right]
        \right\}^2 \vert \psi(t)\rangle.
\end{equation}
\end{widetext}

In the limit considered in this paper we can replace the frequencies 
$\omega_k$ under the radical with the constant $\omega_1^{\rm (at)}$.
After expanding the state vector as in Eq.~(\ref{gen_psi_eq})
and evaluating the sums, the expectation value can be written in terms 
of the amplitudes $b_k(t)$ to find the photon in the various cavity
modes:
\begin{equation} 
\langle \hat{\cal{E}}^2 \rangle 
        = 2 \left(\frac{\hbar\omega^{\rm (at)}_1}{\epsilon_0 V}\right)
          \left\vert \sum_k b_k(t)\sin\left[(k_0 + k)\frac{\pi z}{L}\right]
                        \right\vert^2.
\label{e2_expect_eq}
\end{equation}
(In this expression we have dropped the infinite term arising from 
vacuum expectation value of $\hat{{\cal E}}^2$.)
Evaluation of $\langle \hat{\cal{E}}^2\rangle$ gives a space- and 
time-dependent representation of the localization of the energy
of the photon \cite{BUZ99,LIG02,PUR02}.

The expression for $\langle \hat{\cal{E}}^2\rangle$ in
Eq.~(\ref{e2_expect_eq}) is the square of a complex number that is
analogous to the complex analytic signal describing the classical
field.  We label this quantity $\tilde{\cal E}_{\rm q.m}$, i.e.,
\begin{equation}
\tilde{\cal E}_{\rm q.m.} = \sqrt{\frac{2\hbar\omega^{\rm (at)}_1}
                {\epsilon_0 V}}
          \sum_k b_k(t)\sin\left[(k_0 + k)\frac{\pi z}{L}\right].
\end{equation}
We note that the quantity $\tilde{\cal E}_{\rm q.m.}$ is effectively
equivalent to $\langle 0\vert\hat{E}^{(+)}\vert\psi_\gamma \rangle$,
where $\hat{E}^{(+)}$ is the positive frequency part of the field
quantized in terms of traveling wave modes, which has been identified
as ``the `electric field' associated with [a] single photon state'' by
Scully and Zubairy \cite{SCU97}.  In a previous paper \cite{PUR02} we
extended the quantum-classical correspondence embodied in this
quantity to fields that include scattering from two-level atoms, and
here we extend it to include scattering from three-level atoms that
make up a medium that exhibits electromagnetically induced
transparency.

For ease of comparison with previous results for the detector atom,
we calculate the field at the fixed position $z=3 L/4$ in a cavity that
does not contain the detector atom. With no scattering atom present
we find \cite{PUR02}
\begin{equation}
\tilde{\cal E}_{\rm q.m.}^0 = -i\Theta\left(t-\frac{1}{2}\right)
                \sqrt{\frac{\hbar\omega^{\rm (at)}_1\gamma_1}
                {2\epsilon_0 V}}e^{-\frac{\gamma_1}{2}(t-\frac{1}{2})}.
                        \label{analytic_sig_0_eq}
\end{equation}
The energy density passing the point $z=3L/4$ exhibits an abrupt
turn-on (because of the initial conditions we have chosen) followed by
exponential decay \cite{BUZ99,SCU97,LIG02,PUR02}.  With a three-level
scattering atom present at $z=L/2$ we find
\begin{widetext}
\begin{eqnarray}
\tilde{\cal E}_{\rm q.m.}
      &=&-i\Theta\left(t-\frac{1}{2}\right)
	\sqrt{\frac{\hbar\omega^{\rm (at)}_1\gamma_1}{2\epsilon_0 V}}
	\left\{e^{-\frac{\gamma_1}{2}(t-\frac{1}{2})}
	+\frac{\gamma_2(\gamma_1+i2\delta)
		e^{-\frac{\gamma_1}{2}(t-\frac{1}{2})}}
	{\gamma_1^2 -\gamma_1\gamma_2 - 4\delta^2 + \omega_R^2 +i2\delta
				(2\gamma_1-\gamma_2)}
				\right. 	\nonumber \\
	&& \left.
	+ \frac{\gamma_2\left(\gamma_2-\sqrt{\gamma_2^2 - 4\omega_R^2}
							\right)
       		e^{[-(\gamma_2-\sqrt{\gamma_2^2-4\omega_R^2})/4+i\delta]
						(t-\frac{1}{2})}}
	{\sqrt{\gamma_2^2 - 4\omega_R^2}\left(2\gamma_1 - \gamma_2 
		+ \sqrt{\gamma_2^2 - 4\omega_R^2} + i 4\delta\right)}
				\right. 	\nonumber \\
	&& \left. 
	+ \frac{\gamma_2\left(\gamma_2+\sqrt{\gamma_2^2 - 4\omega_R^2}
							\right)
       		e^{[-(\gamma_2+\sqrt{\gamma_2^2-4\omega_R^2})/4+i\delta]
						(t-\frac{1}{2})}}
	{\sqrt{\gamma_2^2 - 4\omega_R^2}\left(2\gamma_1 - \gamma_2 
		- \sqrt{\gamma_2^2 - 4\omega_R^2} + i 4\delta\right)}\right\}.
\label{analytic_sig_eq}
\end{eqnarray}
\end{widetext}

The first term in curly brackets is just the previous result with no
scattering atom present; the effect of the scattering is contained in
the remaining terms.  The steady state scattering is contained in the
second term in curly brackets.  This term decays at the slow rate
$\gamma_1$ reflecting the envelope of the incident radiation. We
claimed previously that for large $\gamma_3$ the detector atom
response reflects the field incident upon it, and comparison of
Eq.~(\ref{c3s_eq}), which gives the probability amplitude due to
scattering for the detector atom, with Eq.~(\ref{analytic_sig_eq}),
justifies this claim.  In the limit $\gamma_3\gg \gamma_2, \gamma_1$
the factors ${\cal F}_{20}$ through ${\cal F}_{23}$ that arise in
Eq.~(\ref{c3s_eq}) are proportional to the the coefficients in front
of the corresponding exponential terms in Eq.~(\ref{analytic_sig_eq}).

The quantum pulse of Eq.~(\ref{analytic_sig_0_eq}) has the classical 
analog 
\begin{equation} 
\tilde{\cal E}^{0}_{\rm cl.} = \Theta\left(t-\frac{1}{2}\right)
                C e^{-(\frac{\gamma_1}{2}+i\omega_1)(t-\frac{1}{2})}
                        \label{analytic_sig_0_eq_cl},
\end{equation}
where $C$ is a constant.  If such classical pulses are incident on a
thin slab of material with thickness $\Delta z$, density $N$, and
classical susceptibility $\chi$ given by Eq.~(\ref{chi_eq}), then the
classical transmitted field ${\cal E}_{\rm cl.}$ is identical in form
to Eq.~(\ref{analytic_sig_eq}), except that the terms due to
scattering (i.e., those proportional to $\gamma_2$) have a magnitude
characterized by the small dimensionless factor $N\Delta z\vert
d\vert^2\omega/(\hbar\epsilon_0c\gamma_2)$.  A derivation of this 
result is included in an appendix.

\section{Comparison of semi-classical and quantum mechanical scattering}
\label{sect_comparison}

In the semi-classical model the result of weak scattering of a
monochromatic field is contained in Eq.~(\ref{weak_scattering_eq}).
In the limit of narrow bandwidth probe ($\gamma_1\ll\gamma_2$) and
rapid detector atom response ($\gamma_3\gg\gamma_1,\gamma_2$) our
quantum probability for the detector atom to be excited by the
scattered field is approximated by the first term of
Eq.~(\ref{c3s_eq}),
\begin{eqnarray}
c_3^s(t)&\simeq& \Theta\left(t-\frac{1}{2}\right)
	e^{-\frac{\gamma_1}{2}(t-\frac{1}{2})}{\cal F}_{20} \nonumber \\ 
	&\simeq& -\Theta\left(t-\frac{1}{2}\right)
	e^{-\frac{\gamma_1}{2}(t-\frac{1}{2})} 
	\sqrt{\frac{\gamma_1}{\gamma_3}} \nonumber \\
	&& \times \frac{i \delta\gamma_2/2}{\omega_R^2/4 - \delta^2 
		- i\delta\gamma_2/2} \nonumber \\
	&\simeq& c_3^0(t) \frac{i \delta\gamma_2/2}{\omega_R^2/4 - \delta^2 
		- i\delta\gamma_2/2}
\end{eqnarray}
This is exactly the same functional form as the term due to scattering
in the formula for the classical field,
Eq.~(\ref{weak_scattering_eq}).  In the classical formula the
dimensionless factor $N\Delta z\vert
d\vert^2\omega/(\hbar\epsilon_0c\gamma)$ is assumed to be small.  In
our quantum model the magnitude of the scattering is determined by
$\Omega_2$ (or equivalently $\gamma_2$), which characterizes the
coupling of atom 2 to the probe field.  In our one-dimensional model the
coupling to the incident field and the decay rate of atom 2 are both
completely determined by the single parameter $\Omega_2$, which means
that it is not possible to make the effect of the scattering small
without simultaneously making the linewidth of atom 2 very narrow.
In a fully three-dimensional model the decay rate of atom 2 would be
the result of the atom's coupling to many more modes, and not just
those containing the incident field.  This scattering into other modes
would reduce the scattering in the forward direction (the direction of
the detector) from the amount predicted in our simple model.  If the
forward scattering is reduced by a factor $f$, then a more realistic
expression for the excitation of the detector atom is
\begin{equation}
c_3(t) = c_3^0(t) + f c_3^s(t).
\label{f_eq}
\end{equation}
In this formula $f$ plays a role analogous to the dimensionless factor
$N\Delta z\vert d\vert^2\omega/(\hbar\epsilon_0c\gamma)$ in the formulas 
for the transmitted field derived from semi-classical theory.

\section{Temporal-center-of-gravity delay}
\label{sect_cog_delay}

The group velocity of a classical field has a clear interpretation for
quasi-monochromatic pulses whose central frequency is far from a
region of anomalous dispersion: it is the speed at which the peak of
the modulation envelope travels.  Group delays refer to the delay in
the arrival of the peak of a pulse compared to the time expected for
propagation through a vacuum.  The pulses investigated in this paper
have sharp leading edges, and this lack of a smooth modulation envelope 
means that the results of simple classical theory for quasi-monochromatic
pulses should not be expected to be a sufficient guide to
full understanding.  In this section we will investigate 
``temporal-center-of-gravity'' delays in several classical and 
quantum mechanical quantities.

\begin{figure}[t]
\includegraphics[7.5cm,8.0cm][10cm,23.9cm]{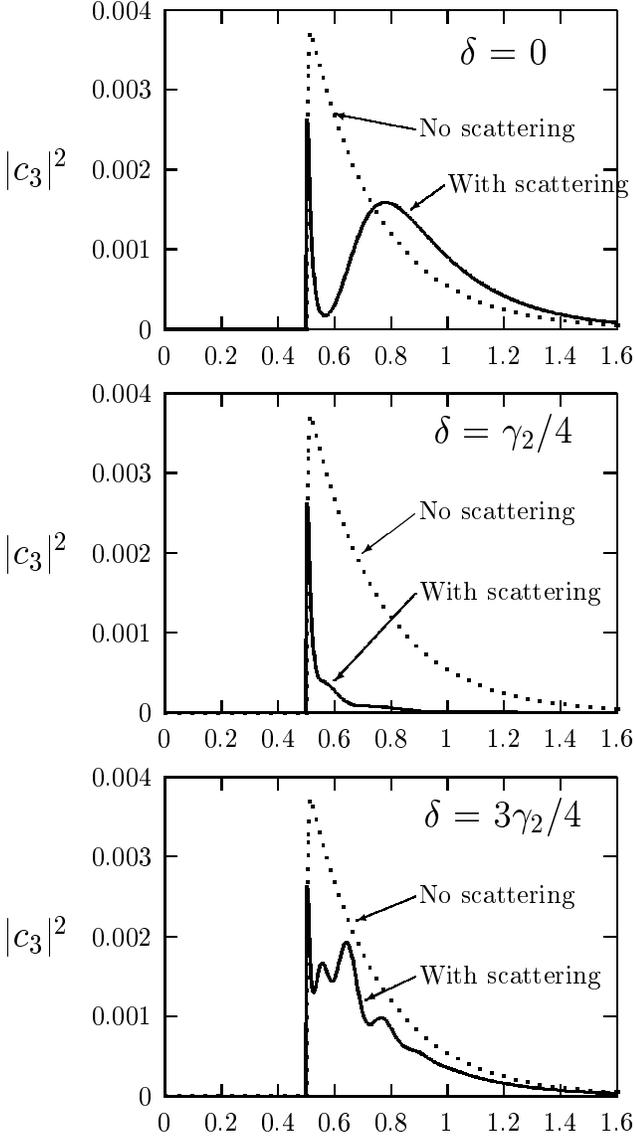}
\caption{Probability for the detector atom to be excited as a function
of time for various values of the detuning $\delta$.  (All effects due
to multiple scattering and reflections have been suppressed.)  The
effect of the ``pulse reshaping'' is dependent on the detuning.  The
temporal center of gravity of the probability is shifted relative to
that in the case of no scattering from relatively late times in the
top graph to earlier times in the middle two graphs, and then back to
later times in the bottom graph.  The decay rates of the atoms are
$\gamma_1 = 4$,$\gamma_2 = 64$, and $\gamma_3 = 1024$; the Rabi
frequency of the B-C transition of atom 2 is $\omega_R = \gamma_2/2 =
32$.}
\label{f_c3}
\end{figure}

The first delay we investigate is derived from $c_3(t)$, the amplitude
for the detector atom to be excited.  As we have argued previously,
this amplitude reflects the strength of the incident field in the
limit that the response time of this atom is very short compared with
other time scales, i.e., $\gamma_3\gg \gamma_1, \gamma_2$.  The effect
of the scattering on this amplitude is evident in Fig.~\ref{f_c3}, in
which $\vert c_3(t)\vert^2$, the probability for the detector atom to be 
excited,  is plotted for three values of the
detuning $\delta$, and also for the case in which no scattering atom
is present.  (The results are plotted for the specific case in 
which $\omega_R = \gamma_2/2.$)
For all detunings, $c_3(t)=0$ for all times earlier than $t=0.5L/c$,
as is expected; all effects on the detector atom occur at times that
preserve causality.  The qualitative shapes of the detector response
depend critically on the detuning $\delta$.  For $\delta=0$ the
steady-state absorption coefficient is zero, and this is reflected in
the fact that at large times there is little effect of the scattering
on the probability.  At shorter times the effect of transient
oscillations have a pronounced affect, and he temporal center of
gravity of the response is clearly shifted toward later times relative
to that of the response with no scattering atom present. The temporal
center of gravity shifts to earlier times as the detuning is
increased, and for values in the vicinity $\delta=\gamma_2/2$ it
occurs earlier than the in the case of no scattering.  This is the
region in which negative values of the group delay occur, as is
illustrated in Fig.~\ref{f_delay}.  The finite response time of the
medium results in large brief transmission of leading edge of the
pulse of the field before the high attenuation of the steady state
sets in.  (A similar effect occurs in scattering from simple two-level
atoms \cite{PUR02}.)  At larger values of the detuning, such as the
case $\delta=\gamma_2$ illustrated in the bottom graph of
Fig.~\ref{f_c3}, most of the probability is again removed at early
times, shifting the temporal center of mass back to later times.

We quantify the ideas illustrated in Fig.~\ref{f_c3} by identifying 
an effective arrival time of the photon with the temporal center of 
gravity of the probability that the detector atom is excited, i.e.,
\begin{equation}
t_{\rm arrival} =  \frac{\int t\vert c_3(t)\vert^2\, dt}
                {\int \vert c_3(t)\vert^2\, dt}.
\end{equation}
(In evaluating the integrals in this equation, we use only the first
term in the series of Eq.~(\ref{c3_gen_eq}), and assume that the decay
rates and distances are such that the effect of multiple scattering is
negligible.)  The delay imposed by the medium is then just the
difference in the arrival times calculated with and without a
scattering atom present,
\begin{equation}
\Delta t_{c_3} = \frac{\int t \vert c_3(t)\vert^2\, dt}
                        {\int \vert c_3(t)\vert^2\, dt}- 
                  \frac{\int t \vert c_3^0(t)\vert^2\, dt}
                        {\int  \vert c_3^0(t)\vert^2\, dt}.
\end{equation}

To explore the effect of weak scattering we rewrite $c_3(t)$ in 
the form of Eq.~(\ref{f_eq}), and assume that $f\ll 1$.  Our quantum
mechanical delay becomes, to first order in $f$,
\begin{eqnarray}
\Delta t_{c_3} &=& \frac{\int t\vert c_3^0(t)+ fc_3^s(t)\vert^2\, dt}
                        {\int \vert c_3^0(t)+ fc_3^s(t)\vert^2\, dt}
- \frac{\int t\vert c_3^0(t)\vert^2\, dt}
        {\int \vert c_3^0(t)\vert^2\, dt} \nonumber \\
        &\simeq& 2f \left[\frac{\int t\, 
                \mbox{Re}\left[c_3^0(t) c_3^s(t)^\ast\right]\, dt}
                {\int  \vert c_3^0(t)\vert^2 dt} \right. \nonumber \\
        && - \left. \frac{\int t\vert c_3^0(t)\vert^2\, dt 
                \int \mbox{Re}\left[c_3^0(t) c_3^s(t)^\ast\right]\, dt}
                        {\left(\int \vert c_3^0(t)\vert^2\, dt\right)^2}
                        \right].
\label{delta_t_qm_eq}
\end{eqnarray}

It is straightforward (but involved) to evaluate the integrals in
Eq.~(\ref{delta_t_qm_eq}) using the expressions for $c_3^0(t)$ and
$c_3^s(t)$ from Eqs.~(\ref{c30_eq}) and (\ref{c3s_eq}).  After 
taking the limit $\gamma_3\rightarrow\infty$ and then letting 
$\gamma_1\rightarrow 0$ we arrive 
at the following expression for the quantum time delay:
\begin{equation}
\Delta t_{c_3} = 4f\gamma_2\frac{(\omega_R^2+4\delta^2)
			\left[(\omega_R^2-4\delta^2)^2 
			- 4\gamma^2\delta^2\right]}		
	{\left[(\omega_R^2-4\delta^2)^2 + 4\gamma^2 \delta^2\right]^2}.
\label{group_delay_qm_eq}
\end{equation}
Recalling that the parameter $f$ characterizes the magnitude of the
scattering in the forward direction in the same way that the quantity
$N\Delta z\vert d\vert^2\omega/ (\hbar\epsilon_0c\gamma)$ does in the
semi-classical case, we see that our ``temporal-center-of-gravity''
delay time for this specific pulse is identical in functional form to
the expression for the group delay of the classical field given in
Eq.~(\ref{group_delay_eq}).  The magnitude of the
temporal-center-of-gravity, however, is twice that given by the group
delay, i.e.,
\begin{equation}
\Delta t_{c_3} = 2 (\Delta t_g)_{\rm classical}.
\end{equation}
The difference in the classical group delay and the
temporal-center-of-mass delay should not be surprising, because the
pulses studied in this paper do not satisfy the quasimonochromatic
condition.  The delay in the arrival time that we have defined is the
result of the reshaping of the ``pulse'' of excitation of the detector
atom. The effect of scattering in the region of resonance is
effectively to redistribute (in time) the probability that the
detector atom will be excited.  The atom can rapidly absorb energy
from the probe field, and the coupling field creates a state of the
system which can then return energy to the probe field at a rate
which can be adjusted, via the Rabi frequency $\omega_R$, to be
arbitrarily slow.  Although we have not demonstrated it explicitly in
this work, we are confident that delays of quasi-monochromatic quantum
pulses with smooth envelopes can be explained in the same manner.

Because $\tilde{\cal E}_{\rm cl.}$, $\tilde{\cal E}_{\rm q.m.}$, and
$c_3(t)$ all have the same functional form (in the large $\gamma_3$
limit), it is easy to see that equivalent delays can be derived from
the classical field using Eq.~(\ref{cl_tcom_delay_eq}), or from the
quantum field using the analogous equation
\begin{eqnarray}
\Delta t_{\langle{\cal E}^2\rangle}&=&
                \left(\frac{\int t \langle\hat{{\cal E}}(z=3L/4)^2\rangle\,
                                        dt}
                {\int \langle \hat{{\cal E}}(z=3L/4)^2\rangle\, dt}
                \right)_{\rm with\, scatterer} \nonumber \\
        &&      -\left(\frac{\int t \langle\hat{{\cal E}}(z=3L/4)^2\rangle\,dt}
                {\int \langle \hat{{\cal E}}(z=3L/4)^2\rangle\, dt}
                \right)_{\rm no\, scatterer}. 
\end{eqnarray}
  
\section{Conclusion}

We have considered a fully quantized model of scattering by
three-level atoms which can exhibit electromagnetically induced
transparency.  Our model is simple enough that we are able to find an
analytical solution describing the complete dynamics of the system.
Using this model we have investigated propagation of single
spontaneously emitted photons through a medium exhibiting
electromagnetically induced transparency, and have defined an
effective time of arrival at a detector atom.  We have also calculated
the expectation value of the field intensity operator, and identified
a quantum analog to the complex analytic signal describing the
classical field.  These quantum mechanical quantities exhibit delays
that are clearly the result of pulse reshaping effects.  We have
compared our delays for quantized one-photon fields with those
calculated for classical fields using the index of refraction
determined from semi-classical theory, and find them to be in
agreement.  

\appendix

\section{Solution using Laplace Transforms}

Before solving the coupled differential equations resulting from the
Schr\"{o}dinger equation, Eqs.~(\ref{c1_schro_eq}-\ref{bk_schro_eq}), we
re-zero the energy scale and measure all energies relative to the
energy of the initial state $\vert e,g,g;\emptyset\rangle$ (ignoring
$\hat{H}_{\rm interaction}$).  Taking the Laplace transform of this
set of coupled differential equations gives a set of coupled algebraic
equations,
\begin{eqnarray}
i\left(s\tilde{c}_1(s)-1\right) &=& \sum_k g_{1k}\tilde{b}_k(s), 
\label{lt1_eq}\\
\nonumber \\
(is+\delta)\tilde{c}_2(s) &=& \sum_k g_{2k}\tilde{b}_k(s) 
		+ \omega_R\tilde{d}(s)/2, 
\label{lt2_eq}\\
\nonumber \\
is\tilde{c}_3(s) &=& \sum_k g_{3k}\tilde{b}_k(s), 
\label{lt3_eq}\\
\nonumber \\
is\tilde{d}(s) &=& \omega_R\tilde{c}_2(s)/2 -\delta\tilde{d}(s),
\label{ltd_eq}\\
\nonumber \\
is\tilde{b}_k(s) &=& \sum_j g^\ast_{jk}\tilde{c}_j(s)+ k\Delta \tilde{b}_k(s) .
\label{ltb_eq}
\end{eqnarray}
We solve these algebraic equations for the quantities
$\tilde{c}_j(s)$, $\tilde{d}(s)$, and $\tilde{b}_k(s)$, and then
perform an inverse Laplace transform to recover the time dependence of
$c_j(t)$, $d(t)$, and $b_k(t)$.  The details of carrying out such
calculations are quite involved, and were completed with the aid of a
computer algebra system \cite{LIG02a}.  In this appendix we outline
our approach and present some of our intermediate results.

We begin by solving Eq.~(\ref{ltd_eq}) for $\tilde{d}(s)$ and
Eq.~(\ref{ltb_eq}) for $\tilde{b}_k(s)$, and substitute the results in
the first three equations, giving
\begin{eqnarray}
s\tilde{c}_1 -1 &=& -i\Delta\left(f_{11}\tilde{c}_1+f_{12}\tilde{c}_2 +
                   f_{11}\tilde{c}_3\right), \label{ltp1_eq} \\
\nonumber \\
\tilde{c}_2 &=& 
		-i\Delta\left(f_{12}\tilde{c}_1 + 
                  f_{22}\tilde{c}_2 +f_{23}\tilde{c}_3\right) \nonumber \\
	&& \times
	\left[\frac{4(s - i\delta)}{4(s - i\delta)^2+\omega_R^2}\right],
				\label{ltp2_eq}\\
\nonumber \\
s \tilde{c}_3 &=& -i\Delta\left(f_{13}\tilde{c}_1 + 
                  f_{23}\tilde{c}_2 + f_{33}\tilde{c}_3\right),
                                \label{ltp3_eq} 
\end{eqnarray}
in which we have defined the dimensionless sums
\begin{equation}
f_{lm} = \frac{1}{\Delta^2}\sum_k\frac{g_{lk}g_{mk}^\ast}
        {\frac{is}{\Delta} - k}.
\end{equation}

In the limit in which the atomic resonance frequencies are very much
greater than the fundamental frequency of the cavity, i.e., 
$\omega_j^{\rm (at)}\gg\Delta$, these
sums may be approximated by extending the range for $k$ from $-\infty$
to $+\infty$, in which case the sums have relatively simple representations
in terms of trigonometric functions.  Explicit expressions for these
sums are given in the Appendix of Ref.~\cite{PUR02}.  

After solving Eqs.~(\ref{ltp1_eq}-\ref{ltp3_eq}) for the quantities
$\tilde{c}_j(s)$ in terms of the sums $f_{jk}$, we rewrite the
hyperbolic trigonometric functions resulting from the sums in terms of
exponential functions, and expand the resulting expressions in powers
of $\exp(-s/2)$ or $exp(-s/4)$.  We also let $c/L=1$ at this point in
the calculation.  The time dependence of the system is recovered by a
term-by-term inverse Laplace transform of the expansion.  The step
function turn-on of the resulting time dependence arises because of
the factors $\exp(-ns/4)$ in the expansion.  The lowest order terms in
our expansions of the Laplace transforms are given here:
\begin{widetext}
\begin{eqnarray}
\tilde{c}_1(s) &=& \frac{2}{2s + \gamma_1} + 
                \frac{2e^{-s/2}\gamma_1\left[ 4s^2+4s\gamma_2 
		-4\delta^2 + \omega_R^2 -4i\delta(2s+ \gamma_2) \right]}
                {(2s+\gamma_1)^2\left[4s^2 + 2s(\gamma_2 - i4\delta) 
		- i2\gamma_2\delta - 4\delta^2 + \omega^2_R\right]} + \cdots \\
\nonumber \\
\tilde{c}_2(s) &=& -\frac{4e^{-s/4}\sqrt{\gamma_1\gamma_2}(s-i\delta)}
                   {(2s+\gamma_1)\left[4s^2 + 2s\gamma_2 - 4\delta^2 + 
			\omega_R^2- i 2\delta(4s+\gamma_2) \right]} 
                                + \cdots \\
\nonumber \\
\tilde{c}_3(s) &=& -\frac{2e^{-s/2}\sqrt{\gamma_1\gamma_3}(
			4s^2 - 4\delta^2 +\omega_R^2 -i 8 s\delta)}
                        {(2s + \gamma_1)(2s+\gamma_3)
                        \left[
			4s^2 + 2s\gamma_2  -4\delta^2+ \omega_R^2 
			- i2\delta(4s + \gamma_2)
			\right]} + \cdots  \\
\nonumber \\
\tilde{d}(s) &=& \frac{i2e^{-s/4}\sqrt{\gamma_1\gamma_2}\omega_R}
                {(2s+\gamma_1)
		\left[4s^2 + 2s\gamma_2 -4\delta^2+ \omega_R^2
		-i 2\delta(4s+\gamma_2)\right]} + \cdots \\
\nonumber \\
\tilde{b}_k(s) &=& -\frac{i2g_{1k}}{(s+ik\pi)(2s + \gamma_1)} + 
                \frac{i4e^{-s/4}\sqrt{\gamma_1\gamma_2}(s-i\delta)g_{2k}}
                {(s+ik\pi)(2s+\gamma_1)
		\left[4s^2 + 2s\gamma_2 -4\delta^2+ \omega_R^2 
		- i2\delta(4s + \gamma_2) \right]} + \cdots 
\end{eqnarray}
The inverse Laplace transform of these expressions gives Eqs.~(\ref{c1_gen_eq}
--\ref{bk_gen_eq}).

\section{Coefficients in Analytical Solution}
The coefficients that appear in Eqs.~(\ref{c1_gen_eq}-\ref{bk_gen_eq})
are given by the following expressions:
\begin{eqnarray}
{\cal F}_1 &=& \frac{\gamma_1\gamma_2\left(-\gamma_1^2+4\delta^2+\omega_R^2
			- i 4\gamma_1\delta\right)}
		    {\left[\gamma_1^2-\gamma_1\gamma_2-4\delta^2 + \omega_R^2
			+i2\delta(2\gamma_1-\gamma_2)\right]} \\
\nonumber \\
{\cal F}_2 &=& \frac{\gamma_1\left[\gamma_1^2 - \gamma_1\gamma_2 -4\delta^2
		+\omega_R^2+i2\delta(2\gamma_1-\gamma_2)\right]
		\left[\gamma_1^2 - 2\gamma_1\gamma_2 -4\delta^2+\omega_R^2
		+i4\delta(\gamma_1-\gamma_2)\right]}
		    {2\left[\gamma_1^2-\gamma_1\gamma_2-4\delta^2 + \omega_R^2
			+i\delta(2\gamma_1-\gamma_2)\right]} \\
\nonumber \\
{\cal F}_3 &=& \frac{2\gamma_1\gamma_2(-\gamma_2+\sqrt{\gamma_2^2-
							4\omega_R^2})}
		{\sqrt{\gamma_2^2-4\omega_R^2}\left(2\gamma_1 - \gamma_2 
			+ \sqrt{\gamma_2^2-4\omega_R^2}+i4\delta\right)^2}\\
\nonumber \\
{\cal F}_4 &=& \frac{2\gamma_1\gamma_2(\gamma_2+\sqrt{\gamma_2^2-4\omega_R^2})}
		{\sqrt{\gamma_2^2-4\omega_R^2}\left(-2\gamma_1 + \gamma_2 
			+ \sqrt{\gamma_2^2-4\omega_R^2}-i4\delta\right)^2}\\
\nonumber \\
{\cal F}_5 &=& \frac{\sqrt{\gamma_1\gamma_2}(\gamma_1 + i 2\delta)}
		{\gamma_1^2 - \gamma_1\gamma_2 - 4\delta^2 +\omega_R^2 
			+i2\delta(2\gamma_1-\gamma_2)}\\
\nonumber \\
{\cal F}_6 &=& \frac{\sqrt{\gamma_1\gamma_2}\left(\gamma_2 - \sqrt{\gamma_2^2
				- 4\omega_R^2}\right)}
		{\sqrt{\gamma_2^2-4\omega_R^2}(2\gamma_1 - \gamma_2 
			+ \sqrt{\gamma_2^2-4\omega_R^2}+i4\delta)}\\
\nonumber \\
{\cal F}_7 &=& \frac{\sqrt{\gamma_1\gamma_2}\left(\gamma_2 + \sqrt{\gamma_2^2
				- 4\omega_R^2}\right)}
		{\sqrt{\gamma_2^2-4\omega_R^2}(-2\gamma_1 + \gamma_2 
			+ \sqrt{\gamma_2^2-4\omega_R^2}-i4\delta)}\\
\nonumber \\
{\cal F}_8 &=&\frac{4\sqrt{\gamma_1\gamma_3}
			(\gamma_1^2-4\delta^2+\omega_R^2+i4\gamma_1\delta)}
		{(\gamma_3-\gamma_1)\left(2\gamma_1-\gamma_2 +
			\sqrt{\gamma_2^2 - 4\omega_R^2}+i4\delta\right)
		\left(-2\gamma_1+\gamma_2 +
			\sqrt{\gamma_2^2 - 4\omega_R^2}-i4\delta\right)}\\
\nonumber \\
{\cal F}_9 &=& -\frac{2\sqrt{\gamma_1\gamma_3}\gamma_2
			\left(\gamma_2-\sqrt{\gamma_2^2-4\omega_R^2}\right)}
		{\sqrt{\gamma_2^2-4\omega_R^2}\left(2\gamma_1-\gamma_2 +
			\sqrt{\gamma_2^2 - 4\omega_R^2}+i4\delta\right)
		\left(-\gamma_2+2\gamma_3 +
			\sqrt{\gamma_2^2 - 4\omega_R^2}+i4\delta\right)}\\
\nonumber \\
{\cal F}_{10} &=& \frac{2\sqrt{2\gamma_1\gamma_3}\gamma_2
			\left(\gamma_2+\sqrt{\gamma_2^2-4\omega_R^2}\right)}
		{\sqrt{\gamma_2^2-4\omega_R^2}\left(-2\gamma_1+\gamma_2 +
			\sqrt{\gamma_2^2 - 4\omega_R^2}-i4\delta\right)
		\left(\gamma_2-2\gamma_3 +
			\sqrt{\gamma_2^2 - 4\omega_R^2}-i4\delta\right)}\\
\nonumber \\
{\cal F}_{11} &=& -\frac{4\sqrt{\gamma_1\gamma_3}
		\left[\gamma_3+i(2\delta+\omega_R)\right]
		\left[\gamma_3+i(2\delta-\omega_R)\right]}
		{(\gamma_3-\gamma_1)\left(-\gamma_2+2\gamma_3 +
			\sqrt{\gamma_2^2 - 4\omega_R^2}+i4\delta\right)
		\left(\gamma_2-2\gamma_3 +
			\sqrt{\gamma_2^2 - 4\omega_R^2}-i4\delta\right)}\\
\nonumber \\
{\cal F}_{12} &=& i\frac{\sqrt{\gamma_1\gamma_2}\omega_R}
		{\gamma_1^2 -\gamma_1\gamma_2-4\delta^2+\omega_R^2
			+i2\delta(2\gamma_1-\gamma_2)}\\
\nonumber \\
{\cal F}_{13} &=& i\frac{2\sqrt{\gamma_1\gamma_2}\omega_R}
		{\sqrt{\gamma_2^2-4\omega_R^2}\left(2\gamma_1 - \gamma_2 
			+ \sqrt{\gamma_2^2-4\omega_R^2}+i4\delta\right)}\\
\nonumber \\
{\cal F}_{14} &=& i\frac{2\sqrt{\gamma_1\gamma_2}\omega_R}
		{\sqrt{\gamma_2^2-4\omega_R^2}\left(-2\gamma_1 + \gamma_2 
			+ \sqrt{\gamma_2^2-4\omega_R^2}-i4\delta\right)}\\
\nonumber \\
{\cal F}_{15} &=& i\frac{2}{\gamma_1-i2k\pi}\\
\nonumber \\
{\cal F}_{16} &=& -i\frac{8\sqrt{\gamma_1\gamma_2}(\gamma_1+i2\delta)}
		{(\gamma_1-i2k\pi)\left(2\gamma_1 - \gamma_2 
			+ \sqrt{\gamma_2^2-4\omega_R^2}+i4\delta\right)
		\left(-2\gamma_1 + \gamma_2 
			+ \sqrt{\gamma_2^2-4\omega_R^2}-i4\delta\right)}\\
\nonumber \\
{\cal F}_{17} &=& -\frac{16\sqrt{\gamma_1\gamma_2}(\delta+k\pi)}
		{(\gamma_1-i2k\pi)\left[-\gamma_2 + 
		\sqrt{\gamma_2^2-4\omega_R^2}+i4(\delta+k\pi)\right]
		\left[\gamma_2 + \sqrt{\gamma_2^2-4\omega_R^2}
		-i4(\delta+k\pi)\right]}\\
\nonumber \\
{\cal F}_{18} &=& i\frac{4\sqrt{\gamma_1\gamma_2}\left(-\gamma_2 +
	\sqrt{\gamma_2^2-4\omega_R^2}\right)}
	{\sqrt{\gamma_2^2-4\omega_R^2}\left[-\gamma_2 
			+ \sqrt{\gamma_2^2-4\omega_R^2}+i4(\delta+k\pi)\right]
	(2\gamma_1-\gamma_2 +\sqrt{\gamma_2^2-4\omega_R^2} +i4\delta)}\\
\nonumber \\
{\cal F}_{19} &=& i\frac{4\sqrt{\gamma_1\gamma_2}\left(\gamma_2 +
	\sqrt{\gamma_2^2-4\omega_R^2}\right)}
	{\sqrt{\gamma_2^2-4\omega_R^2}\left[\gamma_2 
			+ \sqrt{\gamma_2^2-4\omega_R^2}-i4(\delta+k\pi)\right]
	(-2\gamma_1+\gamma_2 +\sqrt{\gamma_2^2-4\omega_R^2}-i4\delta)}
\end{eqnarray}
\end{widetext}

\section{Classical scattering of exponential pulses}
A general pulse $E(z,t)$ may be written in terms of its Fourier 
transform as 
\begin{equation}
E(z,t) = \frac{1}{\sqrt{2\pi}}\int_{-\infty}^\infty A(\omega)
		e^{i(k(\omega)z - \omega t)}\, d\omega,
\end{equation}
where 
\begin{equation}
A(\omega) = \frac{1}{\sqrt{2\pi}}\int_{-\infty}^\infty E(0,t)
		e^{i\omega t}\, dt.
\end{equation}
We consider an incident pulse like that of
Eq.~(\ref{analytic_sig_0_eq_cl}) which for convenience we consider to
arrive at the origin $z=0$ at $t=0$.  The Fourier transform of this
pulse is
\begin{equation}
A(\omega) = \frac{1}{\sqrt{2\pi}}\frac{1}{\left[\gamma_1/2 - 
			i(\omega-\omega_1)\right]}.
\end{equation}
The effect of weak scattering on a Fourier component of a plane wave
is contained in Eq.~(\ref{weak_scattering_eq}), and we construct the
transmitted pulse to the right of the scattering region from the
original Fourier components appropriately modified according to this equation.
Weak scattering means that the dimensionless parameter $N\Delta z\vert
d\vert^2\omega/(\hbar \epsilon_0c\gamma_2)$ is small, and for
convenience we label this parameter $f$.  The transmitted field is thus
\begin{widetext}
\begin{eqnarray}
E_t(z,t) &=& \frac{1}{\sqrt{2\pi}}\int_{-\infty}^\infty A(\omega)
	e^{i\omega(z/c-t)}\left[1 + if
	\frac{\delta\gamma_2/2}{\omega_R^2/4-\delta^2 - i\delta\gamma_2/2}
		\right]\, d\omega   \nonumber \\
	&=& \frac{C}{2\pi}\int_{-\infty}^\infty 
			\frac{1}{\left[\gamma_1/2 - 
			i(\omega-\omega_1)\right]}
	e^{i\omega(z/c-t)}\left[1 + if
	\frac{(\omega-\omega_2)\gamma_2/2}{\omega_R^2/4-(\omega-\omega_2)^2 
		- i(\omega-\omega_2)\gamma_2/2}\right]\, d\omega \nonumber \\
	&=& E_i(z,t) +  \frac{Cif\gamma_2}{4\pi}\int_{-\infty}^\infty 
			\frac{1}{\left[\gamma_1/2 - 
			i(\omega-\omega_1)\right]}
	e^{i\omega(z/c-t)}\left[
	\frac{(\omega-\omega_2)}{\omega_R^2/4-(\omega-\omega_2)^2 
		- i(\omega-\omega_2)\gamma_2/2}\right]\, d\omega.
\end{eqnarray}
\end{widetext}
The integral in the last line above gives the scattered field $E_s(z,t)$. 
We evaluate the integral using contour integration, and note that all
three poles are in the lower half of the complex plane.  For $(z/c-t)>0$ 
the integration along the real axis is closed in the upper-half plane,
giving a result of zero; for $(z/c-t)<0$ the contour is closed
in the lower half plane encircling the poles.  Evaluating this integral 
gives
\begin{widetext}
\begin{eqnarray}
E_s(z,t) &=&  Cf\gamma_2 e^{-i\omega_1 t}\left\{
		\frac{(\gamma_1 + i2\delta)e^{-\frac{\gamma_1}{2}(t-z/c)}}
		{\left[\gamma_1^2-\gamma_1\gamma_2 -4\delta^2 + \omega_R^2
		+i2\delta(2\gamma_1-\gamma_2)\right]} 
			+ \frac{(\gamma_2 - \sqrt{\gamma_2 - 4\omega_R^2})
	   e^{[-(\gamma_2 - \sqrt{\gamma_2 - 4\omega_R^2})/4+i\delta](t-z/c)}}
		{\sqrt{\gamma_2^2 - 4\omega_R^2}\left(2\gamma_1 - \gamma_2 
		+ \sqrt{\gamma_2^2 - 4\omega_R^2} + i 4\delta\right)} 
				\right. \nonumber \\
	&& +  \left.\frac{(\gamma_2 + \sqrt{\gamma_2 - 4\omega_R^2})
	  e^{[-(\gamma_2 + \sqrt{\gamma_2 - 4\omega_R^2})/4+i\delta](t-z/c)}}
		{\sqrt{\gamma_2^2 - 4\omega_R^2}\left(2\gamma_1 - \gamma_2 
		- \sqrt{\gamma_2^2 - 4\omega_R^2} + i 4\delta\right)} \right\}
\end{eqnarray}
\end{widetext}
This is identical in form to the terms describing the scattered field
in the quantum mechanical expression Eq.~(\ref{analytic_sig_eq}).
(The factor $e^{-i\omega_1 t}$ does not appear in 
Eq.~(\ref{analytic_sig_eq}) because of the zero chosen for the energy 
scale in the quantum calculations.)

\begin{acknowledgments}
One of us (T.P.) would like to acknowledge support from Bucknell
Physics Department Research Experiences for Undergraduates Program
(NSF Grant Number PHY-0097424).
\end{acknowledgments}


\begin{thebibliography}{25}
\expandafter\ifx\csname natexlab\endcsname\relax\def\natexlab#1{#1}\fi
\expandafter\ifx\csname bibnamefont\endcsname\relax
  \def\bibnamefont#1{#1}\fi
\expandafter\ifx\csname bibfnamefont\endcsname\relax
  \def\bibfnamefont#1{#1}\fi
\expandafter\ifx\csname citenamefont\endcsname\relax
  \def\citenamefont#1{#1}\fi
\expandafter\ifx\csname url\endcsname\relax
  \def\url#1{\texttt{#1}}\fi
\expandafter\ifx\csname urlprefix\endcsname\relax\def\urlprefix{URL }\fi
\providecommand{\bibinfo}[2]{#2}
\providecommand{\eprint}[2][]{\url{#2}}

\bibitem[{\citenamefont{Scully}(1991)}]{SCU91}
\bibinfo{author}{\bibfnamefont{M.~O.} \bibnamefont{Scully}},
  \bibinfo{journal}{Phys. Rev. Lett} \textbf{\bibinfo{volume}{67}},
  \bibinfo{pages}{1855} (\bibinfo{year}{1991}).

\bibitem[{\citenamefont{Boller et~al.}(1991)\citenamefont{Boller, Imamo\u{g}lu,
  and Harris}}]{BOL91}
\bibinfo{author}{\bibfnamefont{K.-J.} \bibnamefont{Boller}},
  \bibinfo{author}{\bibfnamefont{A.}~\bibnamefont{Imamo\u{g}lu}},
  \bibnamefont{and} \bibinfo{author}{\bibfnamefont{S.~E.}
  \bibnamefont{Harris}}, \bibinfo{journal}{Phys. Rev. Lett.}
  \textbf{\bibinfo{volume}{66}}, \bibinfo{pages}{2593} (\bibinfo{year}{1991}).

\bibitem[{\citenamefont{Field et~al.}(1991)\citenamefont{Field, Hahn, and
  Harris}}]{FIE91}
\bibinfo{author}{\bibfnamefont{J.~E.} \bibnamefont{Field}},
  \bibinfo{author}{\bibfnamefont{K.~H.} \bibnamefont{Hahn}}, \bibnamefont{and}
  \bibinfo{author}{\bibfnamefont{S.~E.} \bibnamefont{Harris}},
  \bibinfo{journal}{Phys. Rev. Lett.} \textbf{\bibinfo{volume}{67}},
  \bibinfo{pages}{3062} (\bibinfo{year}{1991}).

\bibitem[{\citenamefont{Harris et~al.}(1992)\citenamefont{Harris, Field, and
  Kasapi}}]{HAR92}
\bibinfo{author}{\bibfnamefont{S.~E.} \bibnamefont{Harris}},
  \bibinfo{author}{\bibfnamefont{J.~E.} \bibnamefont{Field}}, \bibnamefont{and}
  \bibinfo{author}{\bibfnamefont{A.}~\bibnamefont{Kasapi}},
  \bibinfo{journal}{Phys. Rev. A} \textbf{\bibinfo{volume}{46}},
  \bibinfo{pages}{R29} (\bibinfo{year}{1992}).

\bibitem[{\citenamefont{Hau et~al.}(1999)\citenamefont{Hau, Harris, Dutton, and
  Behroozi}}]{HAU99}
\bibinfo{author}{\bibfnamefont{L.~V.} \bibnamefont{Hau}},
  \bibinfo{author}{\bibfnamefont{S.~E.} \bibnamefont{Harris}},
  \bibinfo{author}{\bibfnamefont{Z.}~\bibnamefont{Dutton}}, \bibnamefont{and}
  \bibinfo{author}{\bibfnamefont{C.~H.} \bibnamefont{Behroozi}},
  \bibinfo{journal}{Nature} \textbf{\bibinfo{volume}{397}},
  \bibinfo{pages}{594} (\bibinfo{year}{1999}).

\bibitem[{\citenamefont{Kash et~al.}(1999)\citenamefont{Kash, Sautenkov,
  Zibrov, Hollberg, Welch, Lukin, Rostovtsev, Fry, and Scully}}]{KAS99}
\bibinfo{author}{\bibfnamefont{M.~M.} \bibnamefont{Kash}},
  \bibinfo{author}{\bibfnamefont{V.~A.} \bibnamefont{Sautenkov}},
  \bibinfo{author}{\bibfnamefont{A.~S.} \bibnamefont{Zibrov}},
  \bibinfo{author}{\bibfnamefont{L.}~\bibnamefont{Hollberg}},
  \bibinfo{author}{\bibfnamefont{G.~R.} \bibnamefont{Welch}},
  \bibinfo{author}{\bibfnamefont{M.~D.} \bibnamefont{Lukin}},
  \bibinfo{author}{\bibfnamefont{Y.}~\bibnamefont{Rostovtsev}},
  \bibinfo{author}{\bibfnamefont{E.~S.} \bibnamefont{Fry}}, \bibnamefont{and}
  \bibinfo{author}{\bibfnamefont{M.~O.} \bibnamefont{Scully}},
  \bibinfo{journal}{Phys. Rev. Lett.} \textbf{\bibinfo{volume}{82}},
  \bibinfo{pages}{5229} (\bibinfo{year}{1999}).

\bibitem[{\citenamefont{Wang et~al.}(2000)\citenamefont{Wang, Kuzmich, and
  Dogariu}}]{WAN00}
\bibinfo{author}{\bibfnamefont{L.~J.} \bibnamefont{Wang}},
  \bibinfo{author}{\bibfnamefont{A.}~\bibnamefont{Kuzmich}}, \bibnamefont{and}
  \bibinfo{author}{\bibfnamefont{A.}~\bibnamefont{Dogariu}},
  \bibinfo{journal}{Nature} \textbf{\bibinfo{volume}{406}},
  \bibinfo{pages}{277} (\bibinfo{year}{2000}).

\bibitem[{\citenamefont{Dogariu et~al.}(2001)\citenamefont{Dogariu, Kuzmich,
  and Wang}}]{DOG01}
\bibinfo{author}{\bibfnamefont{A.}~\bibnamefont{Dogariu}},
  \bibinfo{author}{\bibfnamefont{A.}~\bibnamefont{Kuzmich}}, \bibnamefont{and}
  \bibinfo{author}{\bibfnamefont{L.~J.} \bibnamefont{Wang}},
  \bibinfo{journal}{Phys. Rev. A} \textbf{\bibinfo{volume}{63}},
  \bibinfo{pages}{053806} (\bibinfo{year}{2001}).

\bibitem[{\citenamefont{Lukin and Imamo\u{g}lu}(2001)}]{LUK01}
\bibinfo{author}{\bibfnamefont{M.}~\bibnamefont{Lukin}} \bibnamefont{and}
  \bibinfo{author}{\bibfnamefont{A.}~\bibnamefont{Imamo\u{g}lu}},
  \bibinfo{journal}{Nature} \textbf{\bibinfo{volume}{413}},
  \bibinfo{pages}{273} (\bibinfo{year}{2001}).

\bibitem[{\citenamefont{Harris}(1997)}]{HAR97}
\bibinfo{author}{\bibfnamefont{S.~E.} \bibnamefont{Harris}},
  \bibinfo{journal}{Phys. Today} \textbf{\bibinfo{volume}{50(7)}},
  \bibinfo{pages}{36} (\bibinfo{year}{1997}).

\bibitem[{\citenamefont{Lukin et~al.}(1999)\citenamefont{Lukin, Yellin, Zibrov,
  and Scully}}]{LUK99}
\bibinfo{author}{\bibfnamefont{M.}~\bibnamefont{Lukin}},
  \bibinfo{author}{\bibfnamefont{S.}~\bibnamefont{Yellin}},
  \bibinfo{author}{\bibfnamefont{A.}~\bibnamefont{Zibrov}}, \bibnamefont{and}
  \bibinfo{author}{\bibfnamefont{M.}~\bibnamefont{Scully}},
  \bibinfo{journal}{Laser Physics} \textbf{\bibinfo{volume}{9}},
  \bibinfo{pages}{759} (\bibinfo{year}{1999}).

\bibitem[{\citenamefont{Scully and Zubairy}(1997)}]{SCU97}
\bibinfo{author}{\bibfnamefont{M.~O.} \bibnamefont{Scully}} \bibnamefont{and}
  \bibinfo{author}{\bibfnamefont{M.~S.} \bibnamefont{Zubairy}},
  \emph{\bibinfo{title}{Quantum Optics}} (\bibinfo{publisher}{Cambridge
  University Press}, \bibinfo{address}{Cambridge}, \bibinfo{year}{1997}).

\bibitem[{\citenamefont{Feynman et~al.}(1963)\citenamefont{Feynman, Leighton,
  and Sands}}]{FEY63}
\bibinfo{author}{\bibfnamefont{R.~P.} \bibnamefont{Feynman}},
  \bibinfo{author}{\bibfnamefont{R.~B.} \bibnamefont{Leighton}},
  \bibnamefont{and} \bibinfo{author}{\bibfnamefont{M.}~\bibnamefont{Sands}},
  \emph{\bibinfo{title}{The Feynman Lectures on Physics}}
  (\bibinfo{publisher}{Addison-Wesley}, \bibinfo{address}{Reading, MA},
  \bibinfo{year}{1963}), vol.~\bibinfo{volume}{I}, chap. \bibinfo{chapter}{31
  and 32}.

\bibitem[{\citenamefont{Purdy et~al.}()\citenamefont{Purdy, Taylor, and
  Ligare}}]{PUR02}
\bibinfo{author}{\bibfnamefont{T.}~\bibnamefont{Purdy}},
  \bibinfo{author}{\bibfnamefont{D.~F.} \bibnamefont{Taylor}},
  \bibnamefont{and} \bibinfo{author}{\bibfnamefont{M.}~\bibnamefont{Ligare}},
  \bibinfo{note}{quant-ph/0204009}.

\bibitem[{\citenamefont{Ligare and Oliveri}(2002)}]{LIG02}
\bibinfo{author}{\bibfnamefont{M.}~\bibnamefont{Ligare}} \bibnamefont{and}
  \bibinfo{author}{\bibfnamefont{R.}~\bibnamefont{Oliveri}},
  \bibinfo{journal}{Am. J. Phys.} \textbf{\bibinfo{volume}{70}},
  \bibinfo{pages}{58} (\bibinfo{year}{2002}).

\bibitem[{\citenamefont{Bu\v{z}ek et~al.}(1999)\citenamefont{Bu\v{z}ek,
  Drobn\'{y}, Kim, Havukainen, and Knight}}]{BUZ99}
\bibinfo{author}{\bibfnamefont{V.}~\bibnamefont{Bu\v{z}ek}},
  \bibinfo{author}{\bibfnamefont{G.}~\bibnamefont{Drobn\'{y}}},
  \bibinfo{author}{\bibfnamefont{M.~G.} \bibnamefont{Kim}},
  \bibinfo{author}{\bibfnamefont{M.}~\bibnamefont{Havukainen}},
  \bibnamefont{and} \bibinfo{author}{\bibfnamefont{P.~L.}
  \bibnamefont{Knight}}, \bibinfo{journal}{Phys. Rev. A}
  \textbf{\bibinfo{volume}{60}}, \bibinfo{pages}{582} (\bibinfo{year}{1999}).

\bibitem[{\citenamefont{Ligare and Taylor}(2001)}]{LIG01a}
\bibinfo{author}{\bibfnamefont{M.}~\bibnamefont{Ligare}} \bibnamefont{and}
  \bibinfo{author}{\bibfnamefont{D.~F.} \bibnamefont{Taylor}}
  (\bibinfo{year}{2001}), \bibinfo{note}{paper presented at the Eighth
  Rochester Conference on Coherence and Quantum Optics},
  \urlprefix\url{http://www.eg.bucknell.edu/physics/ligare.html/}.

\bibitem[{\citenamefont{Taylor}(2001)}]{TAY01}
\bibinfo{author}{\bibfnamefont{D.~F.} \bibnamefont{Taylor}},
  \bibinfo{type}{Undergraduate Honors Thesis}, \bibinfo{institution}{Bucknell
  University} (\bibinfo{year}{2001}).

\bibitem[{\citenamefont{Smith}(1970)}]{SMI70}
\bibinfo{author}{\bibfnamefont{R.~L.} \bibnamefont{Smith}},
  \bibinfo{journal}{Am. J. Phys.} \textbf{\bibinfo{volume}{38}},
  \bibinfo{pages}{978} (\bibinfo{year}{1970}).

\bibitem[{\citenamefont{Bloch}(1977)}]{BLO77}
\bibinfo{author}{\bibfnamefont{S.~C.} \bibnamefont{Bloch}},
  \bibinfo{journal}{Am. J. Phys.} \textbf{\bibinfo{volume}{45}},
  \bibinfo{pages}{538} (\bibinfo{year}{1977}).

\bibitem[{\citenamefont{Stey and Gibberd}(1972)}]{STE72}
\bibinfo{author}{\bibfnamefont{G.~C.} \bibnamefont{Stey}} \bibnamefont{and}
  \bibinfo{author}{\bibfnamefont{R.~W.} \bibnamefont{Gibberd}},
  \bibinfo{journal}{Physica} \textbf{\bibinfo{volume}{60}}, \bibinfo{pages}{1}
  (\bibinfo{year}{1972}).

\bibitem[{\citenamefont{Milonni and Knight}(1974)}]{MIL74}
\bibinfo{author}{\bibfnamefont{P.~W.} \bibnamefont{Milonni}} \bibnamefont{and}
  \bibinfo{author}{\bibfnamefont{P.~L.} \bibnamefont{Knight}},
  \bibinfo{journal}{Phys. Rev. A} \textbf{\bibinfo{volume}{10}},
  \bibinfo{pages}{1096} (\bibinfo{year}{1974}).

\bibitem[{\citenamefont{Milonni and Knight}(1975)}]{MIL75}
\bibinfo{author}{\bibfnamefont{P.~W.} \bibnamefont{Milonni}} \bibnamefont{and}
  \bibinfo{author}{\bibfnamefont{P.~L.} \bibnamefont{Knight}},
  \bibinfo{journal}{Phys. Rev. A} \textbf{\bibinfo{volume}{11}},
  \bibinfo{pages}{1090} (\bibinfo{year}{1975}).

\bibitem[{\citenamefont{Meystre and Sargent}(1999)}]{MEY99}
\bibinfo{author}{\bibfnamefont{P.}~\bibnamefont{Meystre}} \bibnamefont{and}
  \bibinfo{author}{\bibfnamefont{M.}~\bibnamefont{Sargent}},
  \emph{\bibinfo{title}{Elements of Quantum Optics}}
  (\bibinfo{publisher}{Springer}, \bibinfo{address}{Berlin},
  \bibinfo{year}{1999}).

\bibitem[{LIG()}]{LIG02a}
\bibinfo{note}{Mathematica notebooks used to perform the calculations are
  available from the authors.}

\end{thebibliography}

\end{document}